\documentclass[article,a4paper,twocolumn,showpacs,superscriptaddress,amssymb]{revtex4}
\usepackage{bm}
\usepackage{amssymb}
\usepackage{amsfonts}
\usepackage{amsmath}
\usepackage{graphicx}% \usepackage[notcite,notref]{tshowkeys}
\usepackage[dvips]{color} % this is required for text in color
\usepackage{epstopdf} %converting to PDF

\begin{document}

\title{Kelvin-Helmholtz instability in $^3$He superfluids in zero temperature limit\protect\footnote{Dedicated to the memory of Alexander Andreev. The physical description of a fermionic quantum condensate, like superfluid $^3$He in the zero temperature limit, rests heavily on his work.}}

\author{V.B.~Eltsov, J.J.~Hosio, M.~Krusius$^\ddag$}     %, G.E. Volovik^{1,2}$}

\address{Department of Applied Physics, Aalto University, P.O. Box 15100, FI-00076 AALTO, Finland}
%$^2$L.D. Landau Institute for Theoretical Physics, Kosygina 2, 119334 Moscow, Russia}

\date{\today}
%\tableofcontents
\vspace{6mm}
\pacs{67.30hp, 47.20.Ft, 03.75.Kk, 97.60.Jd, 26.60-c}
\keywords{superfluid $^3$He, hydrodynamic instability, superfluid phase boundary, vortex formation, superfluid turbulence, zero-temperature limit}

\begin{abstract}

In rotating $^3$He superfluids the Kelvin-Helmholtz (KH) instability of the AB interface has been found to follow the theoretical model above $0.4 \, T_\mathrm{c}$. A deviation from this dependence has been assumed possible at the lowest temperatures. Our NMR and thermal bolometer measurements down to $0.2 \, T_\mathrm{c}$ show that the critical KH rotation velocity follows the extrapolation from higher temperatures. We interpret this to mean that the KH instability is a bulk phenomenon and is not compromised by interactions with the wall of the rotating container, although weak pinning of the interface to the wall is observed during slow sweeping of the magnetic field. The KH measurement provides the only so far existing determination of the interfacial AB surface tension as a function of pressure and temperature down to $0.2 \, T_\mathrm{c}$.

%\vspace{3mm}
%\textbf{PACS numbers:} 67.30hp, 47.20.Ft, 03.75.Kk, 97.60.Jd, 26.60-c

% \textbf{Key words:} superfluid $^3$He, superfluid phase boundary, hydrodynamic instability, vortex formation, superfluid turbulence, zero-temperature limit, interfacial surface tension

\end{abstract}
%\tableofcontents
\maketitle

\section{Kelvin-Helmholtz instability}
\subsection{Introduction}
\label{Intro}

The Kelvin-Helmholtz (KH) instability is a well-known phenomenon of classical hydrodynamics.  The instability condition was derived by Lord Kelvin and Hermann von Helmholtz in the late 1800s. The instability takes place at the interface separating two fluid layers, with flow velocities $\mathbf{v}_1$ and $\mathbf{v}_2$, when the relative flow velocity $\mid \mathbf{v}_1 - \mathbf{v}_2 \mid_\parallel$ parallel to the interface reaches a critical value \cite{HydroDynamics}. Examples of the Kelvin-Helmholtz instability are pervasive in nature whenever flow occurs in layered structures, ranging from micro scale \cite{MicroScale} to intergalactic space \cite{InterGalactic}.

%In the late 1800s the condition for the instability was derived by Lord Kelvin and Hermann von Helmholtz for two inviscid fluid layers (1) and (2), with their relative flow velocity $\mid v_1 - v_2 \mid$,
%\begin{equation}
% {\rho_1 \rho_2 \over \rho_1+\rho_2} (v_1-v_2)^2=2\sqrt{\sigma F},
%\label{InstabilityCriterion}
%\end{equation}
%where $\rho_\mathrm{i}$ are the fluid densities in the two layers, $\sigma$ is the interfacial surface tension, and $F$ the restoring force, usually the gravitational force $F = g ( \rho_2 - \rho_1 )$. At the instability the interface is transformed from a flat to a wavy boundary, as an interfacial capillary wave is formed with the wave vector $k_0=\sqrt{F/\sigma}$.

The superfluid counterpart of the KH instability \cite{Korshunov} was confirmed while the interface between the A and B phases of superfluid $^3$He was investigated in a rotating cylinder \cite{KH-Instability}. In the axial field of a solenoidal magnet, a cylindrical superfluid $^3$He sample is divided by the gradient in the field in A and B phase sections separated by a transverse AB interface. The interface is localized along the contour where the applied field equals the critical value $H_\mathrm{AB} (T,P)$ \cite{Hahn}. Measurements on the stability of the interface in rotation \cite{KH_LongPaper} as a function of rotation velocity $\Omega$, temperature $T$, and liquid pressure $P$ have shown that the instability follows the condition \cite{Volovik}
\begin{equation}
{1\over 2}\rho_{\rm sA}~ (v_{\rm sA}-v_{\rm n}) ^2 +{1\over
2}\rho_{\rm sB}~ (v_{\rm sB}-v_{\rm n}) ^2
    =  \sqrt{\sigma_{\rm{AB}} F_{\rm{AB}}}~.
\end{equation}
\label{SuperfluidInstability}
%\Noindentation
Here $\rho_\mathrm{si}$ are the densities of the superfluid components in the two sections, $\sigma_{\rm{AB}}$ is the interfacial surface tension, and $F_{\rm{AB}}$ the external restoring force acting on the interface owing to the magnetic field gradient: $F_{\rm{AB}}= {1 \over 2} \, (\chi_\mathrm{A} - \chi_\mathrm{B} ) $ $ \mid \!\nabla H^2 \!\mid_{H=H_\mathrm{AB}} \approx (\chi_\mathrm{n} - \chi_\mathrm{B}) \, [H_z \, (dH_z/dz)]_{H=H_\mathrm{AB}}$. The superfluid KH condition differs from the classical criterion for inviscid liquids in that it depends additively on the two counterflow velocities $\mid  \mathbf{v}_{\rm si}-\mathbf{v}_{\rm n} \mid$, \textit{i.e.} the difference of the superflow velocity $\mathbf{v}_{\rm si} $ from its reference value $\mathbf{v}_{\rm n}$. In both phases at constant rotation the normal components are locked to co-rotation with the container and $\mathbf{v}_{\rm n} = \mathbf{\Omega} \times \mathbf{r}$.
%and not on the  difference of the counterflow velocities $\mid  \mathbf{v}_{\rm {sA}}-\mathbf{v}_{\rm {sB}} \mid$.

Initially the experiment is prepared to reside in a metastable state of flow: the A-phase section is in equilibrium rotation, \textit{i.e.} with the superfluid component corotating with the equilibrium number of rectilinear quantized vortices, $\mathbf{v}_{\rm {sA}} \approx \mathbf{v}_{\rm n}$, whereas in the B phase section the superfluid component is vortex-fee and nonrotating, $\mathbf{v}_{\rm {sB}} = 0$. At a critical rotation velocity $\Omega_\mathrm{c} (T,P)$  a burst of vortex loops from the A phase penetrates through the AB interface and introduces vortices into the B phase. The corresponding critical counterflow velocity $\mid \mathbf{v}_{\rm {sB}} - \mathbf{v}_{\rm n} \mid = \Omega_\mathrm{c} \, r$ has been found to agree with Eq.~(1) rather than with the $\sqrt{2}$ times larger classical value of inviscid liquids, for which the reference frame is immaterial and only the relative velocity $\mid {v}_\mathrm{sA} - {v}_\mathrm{sB} \mid $ matters.

\subsection{Extrapolation to very low temperatures}
\label{LowTemp}

The coupling of the superfluid component (and the AB interface) to the external reference frame is provided by the normal component, a cloud of thermal quasiparticle excitations. At very low temperatures the density of this cloud is rarefied to a ballistic gas whose thermal equilibrium is maintained by inelastic collisions with the cylinder walls. We call this regime, where the density of excitations in the bulk liquid follows an exponential temperature dependence, the zero-temperature limit.

In the zero-temperature limit thermal equilibrium between the superfluid and normal components and thereby the validity of Eq.~(1) are on uncertain ground. A number of objections can be raised. Ref.\,\cite{Volovik} suggested that the measured critical velocity might start to depend on the development time of the instability. Or perhaps measurements will approach the higher classical critical velocity of inviscid liquids when the coupling to the external reference frame becomes weaker? Indeed thermal decoupling has been observed in the steady state turbulent motions of the rotating axially propagating vortex front, which is created in the KH instability and thereafter expands along the rotating column, replacing the vortex-free B-phase counterflow ultimately with the equilibrium vortex state \cite{Decoupling}. Energy dissipation in the turbulent vortex front is found to saturate towards the lowest temperatures, whereas an orders of magnitude weaker angular momentum transfer leads to a decoupling in the rotation of the vortex front from the cylinder wall \cite{LowT-Front}.

The influence of the sample cylinder wall (at $r = R$) is a further unknown for the validity of Eq.~(\ref{SuperfluidInstability}). In the rotating cylinder the instability takes place close to the outer circumference at the location where the superfluid counterflow velocity reaches its maximum value.  Measurements above $0.4\,T_\mathrm{c}$ display no evidence that the cylinder wall would directly influence the instability. However, at the lowest temperatures differences in the scattering properties of quasiparticles from diffusely or specularly reflecting surfaces become important and excitations in surface bound states at sub-gap energies give rise to unexpected effects \cite{Lancaster}. Recent measurements with moving objects demonstrate motion decoupled from the higher lying bulk quasiparticle states, which allows supercritical velocities well above the Landau critical velocity $v_\mathrm{L}$ and relaxation effects dominated by the excitations in surface bound state.

%Anomalous and contradictory behaviour has been reported from at least two different measurements on the critical velocity at a moving solid surface: 1) a superconducting NbTi wire in uniform linear motion through a bath of $^3$He-B was not observed to display any type of discontinuity in its velocity properties, even at velocities well above the Landau critical velocity $v_\mathrm{L}$ which normally is the absolute maximum limit for any type of dissipationless motion \cite{Lancaster}, while 2) a micromechanical oscillator consisting of a Si plate which moves in shear-type of motion in the plane of the plate, was observed to have a low critical velocity $\lesssim 0.1\,v_\mathrm{L}$  \cite{Florida}. Thus a deviation of the measured KH velocity from the behaviour predicted by Eq.~(\ref{SuperfluidInstability}) might shed new light also on these questions.

The appearance of B-phase vortices in the KH-instability event is monitored both with NMR and bolometer measurements. The detection is nonlocal and inherently slow since the injected vortex loops have to evolve and expand, \textit{eg.} from the AB interface to the NMR detector coil, before the instability is registered. The evolution from initial tiny vortex loops to the final configuration of rectilinear vortex lines in solid-body rotation is a dynamic process which proceeds in several steps and in the zero-temperature limit requires of order $10^3$ seconds to be completed \cite{KH_LongPaper}. In comparison, the development time of the instability has to be much faster since no measurable delay has been ascribed to it so far.

At the AB interface the order parameter is deflected from the A-phase to the B-phase energy minimum on a length scale of the coherence length $\xi (T,P) \sim 10$ --- 70\,nm. Quasiparticle excitations scatter by the Andreev mechanism \cite{Andreev75Anniversary} from this sharp discontinuity in the order parameter distribution. This leads to a measurable thermal resistance and, in the presence of a heat current along the $^3$He column, to a temperature difference between the two phases. The thermal resistance was measured using a sample arrangement similar to the present one down to $\sim 0.15\,T_\mathrm{c}$ \cite{Haley,MagnonThermometry} and was shown to follow an equilibrium cooling trajectory, \textit{i.e.} the ballistic quasiparticle shower proved to maintain the AB interface in thermal equilibrium with its environment.

Other frequently studied thermal characteristics of the AB interface are its propagation velocity and the associated heating when the phase boundary is in motion. This dissipation is manifested as a friction coefficient $\Gamma$ in the equation of interface motion. It is this finite friction of the AB interface, which in Ref.~\cite{Volovik} leads to Eq.~(1). This is the case even in the limit $\Gamma \rightarrow 0$, since this situation is physically different from the inviscid classical fluid where friction vanishes altogether, $\Gamma \equiv 0$, and a preferred reference frame is irrelevant.

To initiate a successful KH instability event, an initial disturbance of the AB interface has to grow to a corrugation of sufficient depth. This is a complex nonlinear process \cite{Lushnikov} which depends on the relaxation time $\tau_\mathrm{L}$ controlling the response of the interface, while it is damped by orbital viscosity. Measurements on the motion of the interface in an oscillating magnetic field \cite{Arrayas1} shed some light on the question how to estimate the development time of the instability. The dissipation from the interface motion was studied while the  interface was kept in motion with an oscillatory modulation of the confinement field $H_\mathrm{AB}$. At a low temperature of $\sim 0.15\,T_\mathrm{c}$ the measured heating turned out larger than expected, but it could be attributed to orbital viscosity \cite{Arrayas2}. It arises when the orientation of the local orbital order parameter on the B-phase side in the vicinity of the interface is tracking the periodic motion. The orbital relaxation time, which was required to fit the dissipation measured for the periodically driven AB interface, was found to be of order $\tau_\mathrm{L} \sim 0.01$ --- 0.05\,s  \cite{Arrayas2} which is much slower than reported earlier at higher temperatures, but too fast to become measurable with our NMR measuring scheme. Interestingly, orbital viscosity is expected to approach zero at zero temperature \cite{Fisher}, but apparently in the presence of the large magnetic field $H_\mathrm{AB}$ it survives as the dominant source of dissipation.

To clarify the above questions, measurements of the KH instability in the zero-temperature limit are needed. We briefly discuss in Sec.~\ref{Location} the effective radius, at which the KH instability takes place in the rotating cylinder, and then proceed to summarize the main experimental features in Sec.~\ref{ExpPrinciples}. More details can be found in Ref.~\cite{KH_LongPaper}. We  then describe the results on the measured critical rotation velocity $\Omega_\mathrm{c}$ at $0.2\,T_\mathrm{c}$ and 29\,bar liquid pressure in Sec.~\ref{Omega_cResults}. The temporal development after the instability is discussed in Sec.~\ref{KH_DevelopmentTime}. In Sec.~\ref{B->A transition} we compare to the second process which can be engineered to start vortex formation non-invasively in a controlled fashion in the B phase, namely the B $\rightarrow$ A transition during an upward sweep of the magnetic field. Finally, a summary is provided in the concluding Sec.~\ref{Conclusions}.

\section{Characterization of the instability}
\label{Location}

In the rotating equilibrium state the A-phase section houses a regular array of doubly quantized line vortices with an areal density $2 \Omega/\kappa$ (in terms of the single-circulation quantum $\kappa = h/(2m_3)$) so that the total number of double-quantum lines is $N_0 \approx \pi R^2 (\Omega/\kappa)$. At the AB interface the vortex lines curve from perpendicular to parallel to the phase boundary. The radius of curvature is on the order of the inter-vortex distance $\sim 2 r_\mathrm{v}$. On the phase boundary they form a vortex sheet. Within the sheet the double-quantum vortices dissociate \cite{InterfaceVortexSheet} and flare out radially, to meet the cylinder wall in perpendicular orientation. This way the vortex quanta reside at regular intervals in the sheet on the A phase side of the interface and the difference of $\Omega r$ in the velocity of the azimuthally circulating flow across the interface is stabilized.

For the KH instability to happen, a corrugation has to be formed on the AB interface which protrudes to the B-phase side. In the strong counterflow current of the B phase, the vortices covering the corrugation are cut loose into loops which escape into the B-phase section. Consequently the A-phase section remains at all times in the equilibrium vortex state, with $\langle \bm{v}_\mathrm{sA} - \bm{v}_\mathrm{n} \rangle \approx 0$, whereas the B-phase section is initially in the metastable state of vortex-free counterflow, $\bm{v}_\mathrm{sB} - \bm{v}_\mathrm{n} = - \mathbf{\Omega} \times \mathbf{r}$. The critical angular rotation velocity $\Omega_{\mathrm{c}}$ from Eq.~(1) can then be expressed as
\begin{equation}
\Omega_{\mathrm{c}}^2 =  2 \frac {\sqrt{\sigma_\textrm{AB} F_\textrm{AB}}} {\rho_{\rm s,eff} \; R_{\rm eff}^2}\,.
\label{SuperfluidInstability-2}
\end{equation}
Here $\rho_{\rm s,eff}$  is the effective value of the density of the superfluid component in the B phase at the interface. $R_\mathrm{eff}$ is the radial location where the interface loses stability close to the container wall: $R_\mathrm{eff} \lesssim R$.

The instability condition in Eq.~(\ref{SuperfluidInstability-2}) can be approached by slowly increasing $\Omega$. Under the influence of the growing counterflow current, the interface is then probed by perturbations whose amplitude finally becomes comparable to the wave length of the interfacial surface wave $\lambda =  2 \pi \sqrt{\sigma_\textrm{AB} / F_\textrm{AB}}$. A localized corrugation of size $\sim {1 \over 2} \lambda$ eventually develops, which protrudes to the B-phase side.

The bump in the AB interface is an over-damped rapidly decaying corrugation of the interfacial surface wave, whereas the vortex sheet coverage of the interface is moving more slowly owing to mutual-friction damping. The slower moving vorticity is left behind on the B phase side in the form of dislodged vortex loops when the interface springs back to its minimum energy configuration. The vortex loops in the B phase evolve ultimately into rectilinear vortex lines of the B phase section in a process whose nature depends on temperature. As the B phase vortices start to form the counterflow velocity is reduced at the AB interface below the critical value of Eq.~(\ref{SuperfluidInstability-2}) and the phase boundary returns to its sub-critical stable state. The above reconstruction of the critical events in the context of the KH instability in a rotating container is inferred from the analysis of the experimental observations \cite{KH-Review}.

Comparing the size of the initial interface deflection and the density of vortex quanta in the vortex sheet gives an estimate for $R_\mathrm{eff}$. The distance $a$ between two vortex quanta at the rim of the sheet measured along the cylinder wall is the circumference $2\pi R$ divided by $2N_0$, or $a \approx \kappa/(R \Omega)$. The width of the trough is $\sim {1 \over 2}\lambda$ and thus the number of vortex quanta in the trough has to equal $\Delta N = {1 \over 2} \lambda/a$. Experimentally $\Delta N$ has been found \cite{VorInjection} to form a stochastic distribution which peaks at $\Delta N \approx 8$. This gives a numerical estimate for the wave length $ \lambda \approx 0.4$\,mm (using typical experimental values $\kappa = 0.067$\,mm$^2$/s, $R = 3$\,mm, $\Omega \sim 1$\,rad/s), and places the center of the trough at a distance of ${1 \over 4} \lambda \sim 0.1$\,mm from the cylinder wall.

A second consideration constraining the site of the instability event starts from a better estimate of the number of vortices in the A phase section. In the rotating equilibrium state the vortex lines are confined inside a central cluster which is separated by a cylindrical annulus of vortex-free flow from the cylinder wall, to provide the necessary confinement. In Ref.~\cite{Annihilation} the width $d$ of the annulus was measured to be $d \approx 1.6\,r_\mathrm{v}$ for rectilinear doubly-quantized A phase vortex lines, where $r_\mathrm{v} = \sqrt{\kappa/(2\pi \Omega)} \sim 0.15$\,mm (at $\Omega = 1\,$rad/s) is the Wigner-Seitz unit cell radius in the vortex array. The minimum free-energy configuration with the equilibrium number of vortices corresponds to the case where the vortex number $N \lesssim N_0$ obtains its maximum reversible value. If this value is not reached, then $d$ will be correspondingly larger (in order to maintain within the vortex cluster the areal density $2 \Omega/\kappa$ of solid-body rotation).

The actual vortex number is $N \approx N_0 (1-2d/R)$ and the azimuthal flow velocity between the cluster and cylinder wall $v_\mathrm{sA} \approx \Omega r -  \Omega R (1- 2d/R) (R/r)$, i.e. the velocity is falling off with $1/r$ dependence towards the wall. From Eq.~(1) we find that within this distance $d$ from the cylinder wall the stability of the interface is in first order independent of $r$, if one considers an expansion in terms of $d/R$. Therefore the stability is not necessarily lost at maximum $r = R$, but the center of the trough can comfortably fit up to a distance $d$ from the wall. Indeed, all experimental values place $R_\mathrm{eff}$ between 2.5 and 2.7\,mm \cite{KH_LongPaper} (while the directly measured value of the cylinder is $R = 2.93$\,mm). In this work we use $R_\mathrm{eff} = 2.67\,$mm, as determined in Ref.~\cite{KH_LongPaper}. Thus, having the site of the instability at a distance $R - R_\mathrm{eff} \gtrsim  r_\mathrm{v}$ from the wall, we might predict that the instability is not strongly influenced by the container wall.

The instability in the middle of the long rotating column starts a sequence of dynamical vortex events which are laminar above about $0.6\,T_\mathrm{c}$ and dominated by turbulence below. In the turbulent regime the vortex loops, which are released by the KH instability into the vortex-free B phase flow, first interact in a sudden turbulent burst. The resulting turbulent tangle organizes itself into a vortex front \cite{FrontPropagation},  which rotates and propagates longitudinally towards the vortex-free end of the sample cylinder. It is connected to a trailing twisted cluster of vortex lines. At low temperatures, when $T \rightarrow 0$, the number of vortices in the twisted cluster falls increasingly below the solid-body rotation value \cite{KH-Review}. When the front eventually reaches the far end of the cylinder and turbulence ceases, slow laminar evolution towards the equilibrium vortex state takes over: the number of vortices organizes itself close to that corresponding to equilibrium rotation, the twist unwinds, and the vortices straighten to rectilinear lines.

In the following these features of the KH instability are studied in as much as they become important in our measurements at $\sim 0.2\,T_\mathrm{c}$.

%The $\Delta N$ vortices which manage to break through the interface cover a trough of size equal to one half of the wave length of the interfacial perturbation: ${1 \over 2}\lambda \approx 2\pi \sigma / (\rho_\mathrm{s,eff} R^2 \Omega^2)$. Thus the number of these vortices is $\Delta N = {1 \over 2} \lambda/a \approx 2 \pi \sigma / (\rho_\mathrm{s,eff} \kappa R \Omega)$.

\section{Measurement techniques}
\label{ExpPrinciples}

The liquid $^3$He sample is contained in a 110\,mm long fused quartz glass tube with an inner diameter of 5.85\,mm. At its bottom end the sample volume terminates in an end plate with a central orifice of 0.75\,mm in diameter (see Ref.~\cite{KH_LongPaper}, Fig. 1). The orifice is needed to reduce the leakage of vortices from below the orifice into the sample volume. A superconducting solenoid around the quartz cylinder provides an axially oriented magnetic field for stabilizing a layer of A phase in the central section of the sample column.

In the present experiments, a second compartment of 12\,mm height was added below the end plate (see Ref.~\cite{ThermalDetection}, Fig. 1). This volume houses two quartz tuning fork resonators. The tuning forks are included to provide a sensitive temperature reading in the regime of ballistic quasiparticle flight below $0.3\,T_\mathrm{c}$. The tuning forks are mounted on an additional division plate with an even smaller pinhole of 0.3\,mm diameter. The small volume above this second aperture with its temperature sensors forms a bolometer which is used for measuring the thermal signal from turbulent vortex motion in the sample volume, as triggered by the KH instability. Its thermal time constant $\sim 25$ s is defined by the thermal resistance of the 0.3\,mm  orifice and the heat capacity of the $^3$He-B in the bolometer volume. The cooling of all liquid $^3$He above the 0.3\,mm  aperture takes place via thermal contact to a heat exchange sinter located 45\,mm lower on the nuclear refrigeration stage. In rotation the two compartments below the upper 0.75\,mm orifice are typically filled with vortices, while above the 0.75\,mm orifice in the sample volume the situation can be manipulated by the experimenter. We use purified $^3$He gas for our sample with a $^4$He concentration below 50\,ppm.

For setting up the initial metastable state of flow for the KH measurement, temperatures below $0.4\,T_\mathrm{c}$ become challenging owing to the need to avoid the last few dynamic remanent vortices \cite{RemanentVortices}. These left-over vortices are of irregular shape and slide along the smooth container wall, propelled by the flow created by their own curvature. With decreasing temperature mutual friction dissipation in the B phase decreases exponentially and a solitary floating remnant becomes ever more long-lived: for the final few remnants at $\Omega = 0$ their slow motion towards annihilation can take hours, depending on temperature. If $\Omega$ is increased, the remnants become mobile in the applied flow and then suffer the turbulent instability \cite{TurbulentInstability}. This process removes most of the vortex-free superfluid counterflow, filling the B phase section with a large number of rectilinear vortex lines, usually at a rotation velocity well below that of the KH instability: instead of the vortex-free state the situation now approaches the equilibrium state of rotation.

Furthermore, extrinsic vortices may leak through the orifice and start turbulence in the first B phase compartment above the 0.75\,mm orifice, if rotation is increased too rapidly \cite{VortexLeaks}. This is the reason why the bottom B phase section below the A phase layer is more prone to uncontrolled vortex formation. In contrast, the top B phase section is protected by the barrier formed by the A-phase layer and its two interfaces. Overall as a rule, the lower the temperature, the more care one has to invest in the elimination of unwanted vortices.

A useful technique for avoiding remnants turned out to be a warm-up to the laminar regime of vortex motion above $0.6\,T_\mathrm{c}$ where the remnants annihilate rapidly. After a sufficiently long waiting period at $\Omega = 0$, rotation is increased to just below the expected low-temperature KH $ \Omega_\mathrm{c}$. This situation is then maintained for a sufficiently long time, to make sure that vortex-free counterflow persists. Next the sample is cooled in constant rotation to the lowest temperature. To secure a smooth approach of the KH instability, instead of increasing $\Omega$ to $ \Omega_\mathrm{c}$, the current $I_\mathrm{b}$ in the A-phase barrier magnet is swept downward at constant slow rate, until the instability takes place. The reason for preferring a current sweep to acceleration of rotation is to minimize mechanical interference: mechanical stability of the cryostat rotation at constant $\Omega$ is maximized by using preselected $\Omega$ values away from mechanical resonances.

\begin{figure}[tt!]
 \centerline{
  \includegraphics[width=0.97\linewidth]{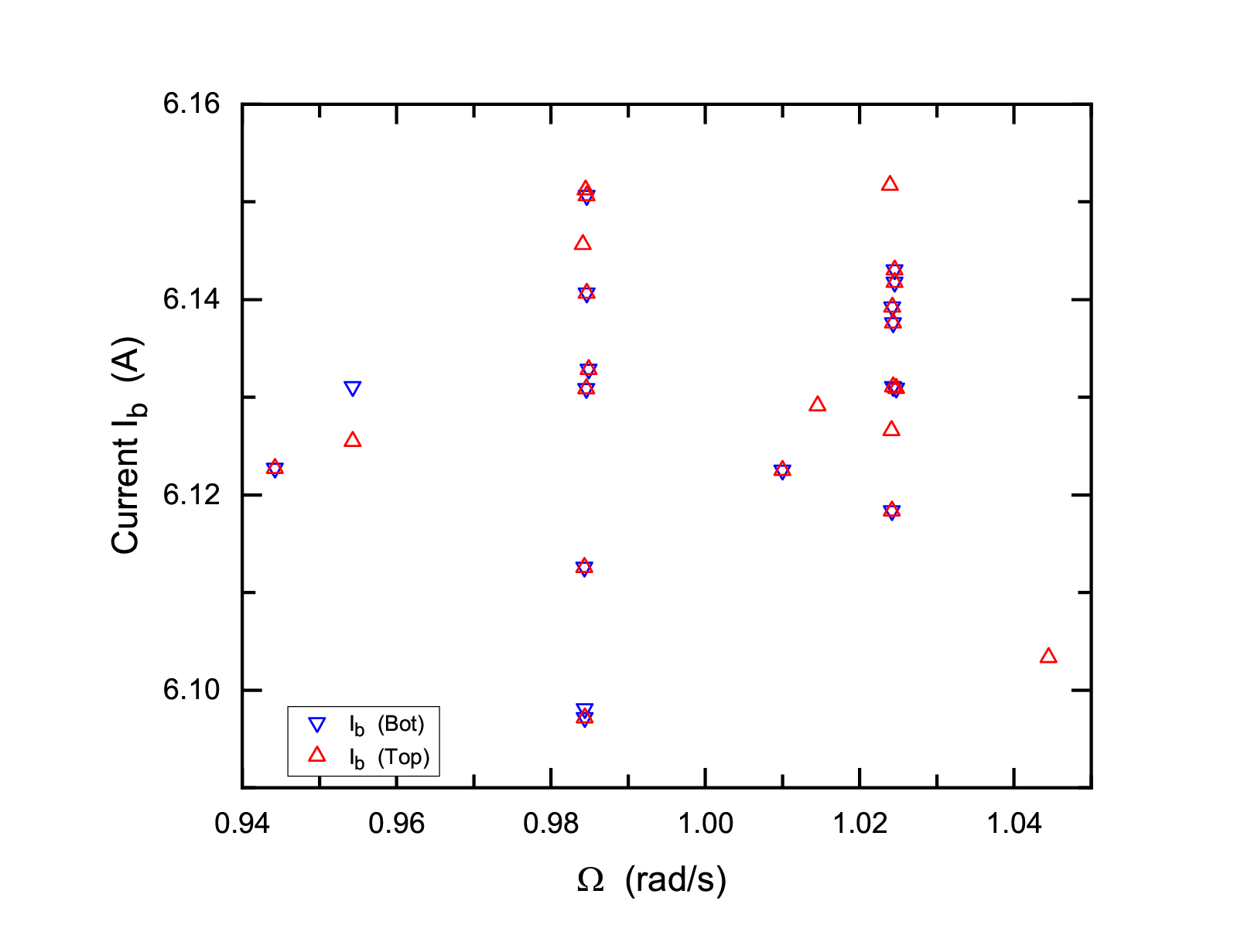}}
\caption{The distribution of the barrier magnet current $I_\mathrm{b}$  at the KH instability in different measurement runs at preselected values of $\Omega$. The plotted $I_\mathrm{b}$ represents its measured value at the moment when the bolometer signals that the dissipation from turbulent vortex motion starts. The vertical spread of the data is interpreted to characterize pinning of the top and bottom AB interfaces at the wall of the sample container. As an example, at $\Omega = 0.984$\,rad/s the average current is 6.134\,A with a standard deviation of 0.019\,A.}
\label{AB-InterfacePinning}
\end{figure}

In earlier measurements where $\Omega$ is increased to trigger the KH instability, the AB interfaces remain largely unchanged until the instability happens. In contrast, when $I_\mathrm{b}$ is swept downward, the two interfaces move closer to each other and the volume of the A-phase layer is reduced. Unfortunately, owing to its surface tension, an AB interface can be expected to be occasionally pinned at surface irregularities and might not follow perfectly smoothly the linear downward moving $I_\mathrm{b}$ sweep. This introduces a new source of scatter in the results. In Fig.\,\ref{AB-InterfacePinning}, 42 data points are displayed from 23 measurement runs at preselected values of $\Omega$ as a function of the $I_\mathrm{b}$ reading when the KH instability takes place. The instability is signalled by the bolometer (see Fig.\,\ref{Bolometer})  when the extra dissipation from turbulent vortex front propagation starts. The random spread in  $I_\mathrm{b}$ values at these different distinct $\Omega$ settings, we interpret as indicative of weak AB interface pinning.

In most cases the KH instabilities in the top and bottom sections occur at nearly the same value of $I_\mathrm{b}$ and a time interval separating the two KH processes is not resolved by the bolometer output. In this case also the NMR alerts signalling the arrival of the turbulent vortex front to the top and bottom NMR detector coils are closely spaced. However, there are examples where the top and bottom events are more widely separated in time and two distinct bolometer signals can be distinguished, particularly if the $I_\mathrm{b}$ sweep is interrupted at the moment when the first instability event is detected. Often it is then the instability in the bottom section which happens first at slightly higher $I_\mathrm{b}$. This might be connected with the fact that dirt and solidified gas particles are more likely to be found on the quartz glass wall towards the open end of the cylinder. Such defects might enhance interface pinning.

In Fig.\,\ref{AB-InterfacePinning}, there are fewer data points for the bottom B phase section  because it had been filled with vortices already at an earlier stage during the measurement,  most likely caused by a vortex leak through the orifice. Also, we note that within the noise limitations of the bolometer signal, we did not encounter a distinct example where the KH instability would have happened while the $I_\mathrm{b}$ sweep was intermittently stopped. We interpret this to mean that the weak-pinning-limit applies to the AB interface in these measurements.

%%%%%%%%%%%%%%%%%%%%%%%%%%%%%%%%%%%%%%%%%%%%%%%%%%%%%%%%%%%%%%%%%%%
%%%%%%%%%%%%%%%%%%%%%%%%%%%%%%%%%%%%%%%%%%%%%%%%%%%%%%%%%%%%%%%%%%%

The above preparation procedure yields reproducible results and a consistent set of values $(T, \Omega, I_\mathrm{b}, P)$ for the KH instability.  The measurement is not obscured by vortex remanency, since remanent vortices would have started developing already at an earlier stage during the preparation procedure. However, an increased scatter may result from AB interface pinning and from initiating the KH instability by sweeping down $I_\mathrm{b}$ might affect the turbulent vortex formation process when compared to if $\Omega$ is increased at stable $I_\mathrm{b}$.

The above recipe for eliminating remanent vortices was checked by increasing $\Omega$ slowly as high as possible in the state of vortex-free counterflow in zero barrier field ($I_\mathrm{b}=0$), to obtain an estimate of the break-down point where vortices start forming spontaneously. At $0.19\,T_\mathrm{c}$ and 29 bar pressure, vortex formation started at 1.54\,rad/s (or 4.5 mm/s). This limit we believe to represent the intrinsic critical flow velocity at the wall of the fused quartz glass tube during the time of the measurements, \textit{i.e.} the counterflow velocity at which vortex formation starts spontaneously at an isolated sharp protuberance on the cylinder wall \cite{VorFormation}.

The present data on the critical KH rotation velocity $\Omega_\mathrm{c} (T, I_\mathrm{b}, P)$ were collected in measurements where the primary objective was to determine the heat release from the turbulent vortex front motion following the KH instability \cite{Decoupling}.  The earlier KH measurements \cite{KH_LongPaper} mapped the instability behavior above $0.4\,T_\mathrm{c}$ using the NMR frequency shift of the so-called ``counterflow NMR absorption peak'' for thermometry. Between these two sets of measurements the rotating nuclear demagnetization cryostat was moved to a different building and was completely refurbished such that the heat leak to the sample container was reduced by more than an order of magnitude. Quartz tuning forks were introduced for more sensitive temperature measurement below $0.4\,T_\mathrm{c}$. This made it possible to observe thermal signals in the low pW regime \cite{Decoupling,ThermalDetection}. As seen in Fig.~\ref{w_exp29bar}, good agreement is obtained between the earlier and later KH results, in spite of the extensive changes in the measuring equipment.

\vspace{3mm}

\section{Temperature dependence of KH instability}
\label{Omega_cResults}

\begin{figure}[tt!]
 \centerline{
  \includegraphics[width=0.97\linewidth]{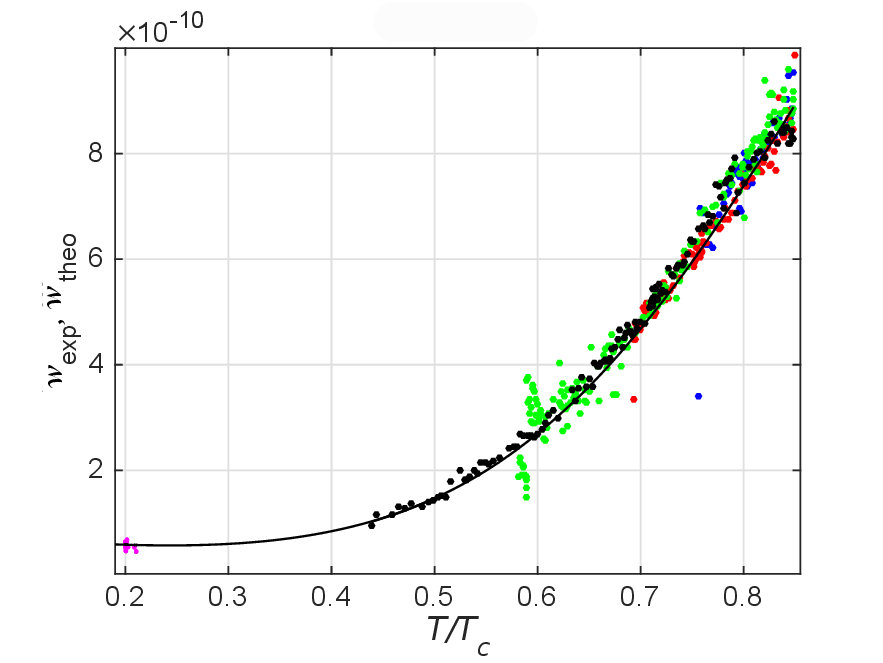}}
\caption{KH instability at different pressures plotted as $w_\mathrm{exp} = (\Omega_\mathrm{c}\, R_\mathrm{eff})^4/(4H_\mathrm{AB}\, (\nabla H)_\mathrm{AB})$ \textit{vs} temperature $T/T_\mathrm{c}$. The solid curve provides a comparison to its fitted theory-based counterpart $w_\mathrm{theo} = \sigma_\mathrm{AB} \, \Delta \chi/\rho^2_\mathrm{s,eff}$. The measurements above $0.4\,T_\mathrm{c}$ are from Ref.~\cite{KH_LongPaper} while the points around $0.2\,T_\mathrm{c}$ are the subject of this report at 29.0\,bar pressure.
}
\label{w_exp29bar}
\end{figure}

An informative comparison of the measured $\Omega_\mathrm{c} (T)$ to Eq.\,(\ref{SuperfluidInstability-2}) is obtained by rewriting the formula in the form
\begin{equation}
w_\mathrm{exp} = \frac{(\Omega_{\mathrm{c}}\, R_\mathrm{eff})^4} {2\,[\nabla H^2]_{H=H_\mathrm{AB}}} =  \frac {\sigma_\mathrm{AB} \; \Delta \chi} {\rho_\mathrm{s,eff}^2} = w_\mathrm{theo} \,,
\label{SuperfluidInstability-4}
\end{equation}
where the left side $w_\mathrm{exp}$ contains experimental and the right side $w_\mathrm{theo}$ theoretical input. In Fig.~\ref{w_exp29bar} data points on $w_\mathrm{exp}$ are plotted as a function of $T/T_\mathrm{c}$ and are compared to a continuous curve representing $w_\mathrm{theo}$. The new data can be seen on the far left in the low-temperature corner at $\sim 0.2 \, T_\mathrm{c}$. The data points above $0.4\,T_\mathrm{c}$ are from Ref.~\cite{KH_LongPaper}.

The surface tension was expressed in Ref.~\cite{KH_LongPaper} in the form  $\sigma_\mathrm{AB} = \sigma_0 \; (1-T/T_\mathrm{c})^{3/2} \; (1- a_\mathrm{H}\,H_{\mathrm{AB}}^2)$. The value of the zero-field parameter $\sigma_0 (P)$, which accounts for the liquid pressure dependence, was adjusted by fitting the high-temperature $\Omega_\mathrm{c} (T)$ data measured at constant pressure $P$ to Eq.~(\ref{SuperfluidInstability-4}). The coefficient of the first order term in the expansion of $\sigma_\mathrm{AB}$ as a function of magnetic field $a_\mathrm{H}(P)$ \cite{Kleinert_AB}, in turn, was fitted using the low-temperature data for $\Omega_\mathrm{c} (T)$.

%%%%%%%%%%%%%%%%%%%%%%%%%%%%%%%%%%%%%%%%%%%%%%%%%%%%%%%%%%%%%%%%%%%
%%%%%%%%%%%%%%%%%%%%%%%%%%%%%%%%%%%%%%%%%%%%%%%%%%%%%%%%%%%%%%%%%%%

For the analysis of the new low-temperature data in Fig.~\ref{w_exp29bar} we have used the earlier value of $\sigma_0 (P)$. The coefficient $a_\mathrm{H}$ is kept as a free parameter; its value is adjusted by fitting each measurement at $\sim 0.2 \, T_\mathrm{c}$, with varying combinations of the experimental variables $\Omega$, $T$, and $I_\mathrm{b}$, to Eq.~(\ref{SuperfluidInstability-2}).  The average of all data gives $a_\mathrm{H} = (2.02 \pm 0.13)$ (in $1/\mathrm{Tesla})^2$, which should be compared with the earlier value $(1.72 \pm 0.12)\; \mathrm{Tesla}^{-2}$ from Ref.~\cite{KH_LongPaper}. Note that the magnetic field dependence amounts at $0.2 \, T_\mathrm{c}$ to a 40\,\% reduction below the extrapolation of a fictive zero-field value owing to this first-order magnetic-field-dependent correction.

\begin{figure}[tt!]
 \centerline{
  \includegraphics[width=0.97\linewidth]{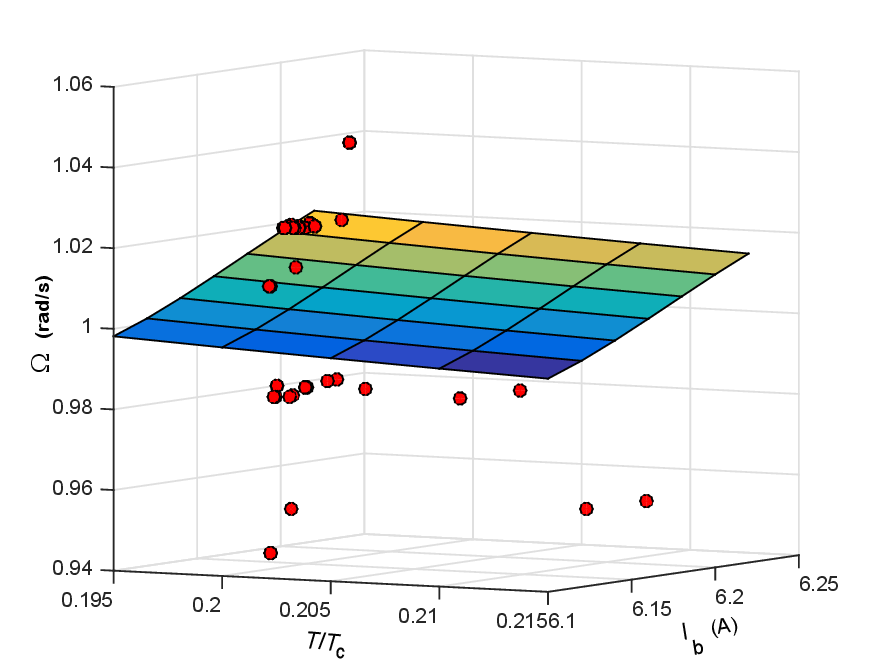}}
\caption{The low-temperature data from Fig.~\ref{w_exp29bar} replotted in a 3-dim coordinate system $(\Omega, T/T_\mathrm{c}, I_\mathrm{b})$. The calculated surface $\Omega_c(T/T_\mathrm{c},I_\mathrm{b})$ provides a comparison to Eq.~(\ref{SuperfluidInstability-2}) when the magnetic-field-dependent surface tension parameter $a_\mathrm{H} = 2.02 $\,T$^{-2}$, \textit{i.e.} the same as used to generate the solid curve in Fig.~\ref{w_exp29bar}. $I_\mathrm{b}$ is here the experimentally scanned variable for triggering the KH instability, with a current-to-field correspondence 6.15 A $\mathrel{\widehat{=}} 0.552$\,T.
}
\label{Omega_c}
\end{figure}

To gain better  oversight with higher resolution, we replot the new low-temperature data in the 3-dim coordinate system of Fig.~\ref{Omega_c} with the variables $(\Omega, T/T_\mathrm{c},I_\mathrm{b})$, together with the surface representing the function $\Omega_\mathrm{c} = f(T/T_\mathrm{c}, I_\mathrm{b})$. This plot illustrates the scatter of the data in terms of the three experimental variables. Most measurement sessions yield a result for $\Omega_\mathrm{c}$ both from the top and bottom sample sections, but since the two sets of data do not show systematic differences, we do not distinguish between them in Fig.~\ref{Omega_c}.

%%%%%%%%%%%%%%%%%%%%%%%%%%%%%%%%%%%%%%%%%%%%%%%%%%%%%%%%%%%%%%%%%%%
%%%%%%%%%%%%%%%%%%%%%%%%%%%%%%%%%%%%%%%%%%%%%%%%%%%%%%%%%%%%%%%%%%%

%As seen from Figs.~\ref{w_exp29bar} -- \ref{Omega_c}, the measurements yield stable and reproducible results for the KH instability. For instance, while sweeping down $I_\mathrm{b}$, the instability occurred in the narrow range $(6.14 \pm 0.02)\,$A, although $\Omega$ and $T$ were not adjusted to always the exactly same values, but included a variation $(1.00 \pm 0.03)\,$ rad/s and $(0.202 \pm 0.03)\,T_\mathrm{c}$, respectively.

As seen in Fig.~\ref{w_exp29bar}, the KH instability follows a predictable trajectory down to the lowest temperatures, obtained as a straightforward extrapolation from higher temperatures.  Also the scatter of the data remains essentially unchanged. Therefore the bulk-volume model of the KH instability continues to apply in the ballistic quasiparticle scattering regime and complications from interactions with the container wall are not seen.

\section{Temporal response after the KH instability}
\label{KH_DevelopmentTime}

The study of the recovery to a new equilibrium state in the long rotating column after the KH instability has brought much new insight in superfluid dynamics. Earlier measurements had shown the recovery to follow a predictable and reproducible path \cite{{FrontPropagation},KH_LongPaper}. The response time was dominated by the flight time of the turbulent vortex front, followed by its trailing vortex cluster, which propagate along the initially vortex-free B phase column. Above $0.4\,T_\mathrm{c}$ the instability could be conveniently triggered with a small step increase of rotation, typically in the amount of $\Delta \Omega \sim 0.02$ --- 0.05\,rad/s, which makes the response time measurement simple.

\begin{figure}[tt!]
 \centerline{
  \includegraphics[width=0.97\linewidth]{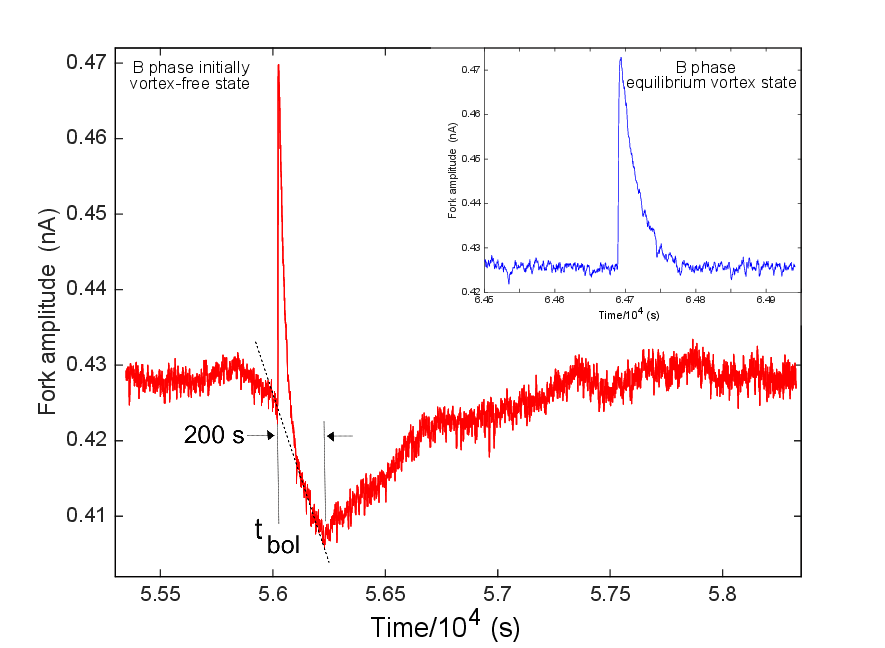}}
  \caption{Two bolometer recordings, while the magnetic field is increased to start the B $\rightarrow$ A transition (see Sec.~\ref{B->A transition}). The oscillation amplitude of a quartz tuning fork resonator is plotted (in arbitrary units) \textit{vs} time. The B $\rightarrow$ A transition is signalled by the sharp cooling spike (at $t_\mathrm{bol}$). This  spike is the only anomaly when the B phase is in  equilibrium rotation (inset). When the B phase is initially in metastable vortex-free rotation (main panel) the peak is followed by heating from the two turbulent vortex fronts. The minimum in signal amplitude, representing the maximum in accumulated heating (at 200 s after the B $\rightarrow$ A transition), marks the moment when the vortex fronts reach the end plates of the rotating column. The later broad shoulder is produced by heating from slow laminar formation of the equilibrium vortex states in both B-phase sections. (Measuring conditions: $T/T_\mathrm{c} = 0.188$, $\Omega = 1.21$ rad/s, $I_\mathrm{b} = 6.01$ A, $P = 29.0$ bar.)
 }
\label{Bolometer}
\end{figure}

At $0.2\,T_\mathrm{c}$, in contrast, when the preferred triggering method is a downward sweep of $I_\mathrm{b}$, the bolometer signal is required to identify the moment when the dissipation from vortex motion starts (see Fig.\,\ref{Bolometer}). Here the front moves slowly and the flight time to the NMR detection coil grows to $\gtrsim 200\,$s. %Also, it turns out that the measured response time displays an unexpected wide spread to longer values.

%%%%%%%%%%%%%%%%%%%%%%%%%%%%%%%%%%%%%%%%%%%%%%%%%%%%%%%%%%%%%%%%%%%
%%%%%%%%%%%%%%%%%%%%%%%%%%%%%%%%%%%%%%%%%%%%%%%%%%%%%%%%%%%%%%%%%%%%

The bolometer signal in Fig.~\ref{Bolometer} is recorded in the presence of a slow $I_\mathrm{b}$ sweep. The sudden steep linear increase in the accumulated heat (dashed line in Fig.~\ref{Bolometer})  arises from the turbulent dissipation while the vortex front propagates along the column. Its extrapolation to $t_\mathrm{bol}$ we equate with the instant when vortex formation starts. The end of the steep increase, the prominent  maximum in the accumulated heat, signals the moment when the vortex fronts reach the top and bottom end plates of the sample cylinder and turbulence subsides. Slightly before this the leading edge of the vortex front passes through the NMR coil creating an identifiable change in the NMR line shape which marks the time point $t_\mathrm{NMR}$ \cite{ThermalDetection}. The measured response time $t_\mathrm{NMR} - t_\mathrm{bol}$, characterized by the constant-slope heat release in  Fig.~\ref{Bolometer}, corresponds thus per its experimental definition roughly to the turbulent propagation. The later broad maximum in the bolometer output is caused by the largely laminar evolution towards the B phase rotating equilibrium state \cite{Decoupling}.

In the example of Fig.~\ref{Bolometer} the vortex fronts in the top and bottom sections move almost in unison which maximizes the bolometer signal. The main source of uncertainty in the measurement of the response time is caused by the noise in the bolometer reading.  The problem are temporal variations in the residual heat leak flowing through the sample cylinder, which appear as bumps and spikes in the background drift of the bolometer signal. Their origin is mechanical noise, partly from the laboratory building, which gets amplified by the mechanical resonances in the cryostat structure rotating at constant $\Omega$.

%%%%%%%%%%%%%%%%%%%%%%%%%%%%%%%%%%%%%%%%%%%%%%%%%%%%%%%%%%%%%%%%%%%
\begin{figure}[tt!]
 \centerline{
  \includegraphics[width=0.97\linewidth]{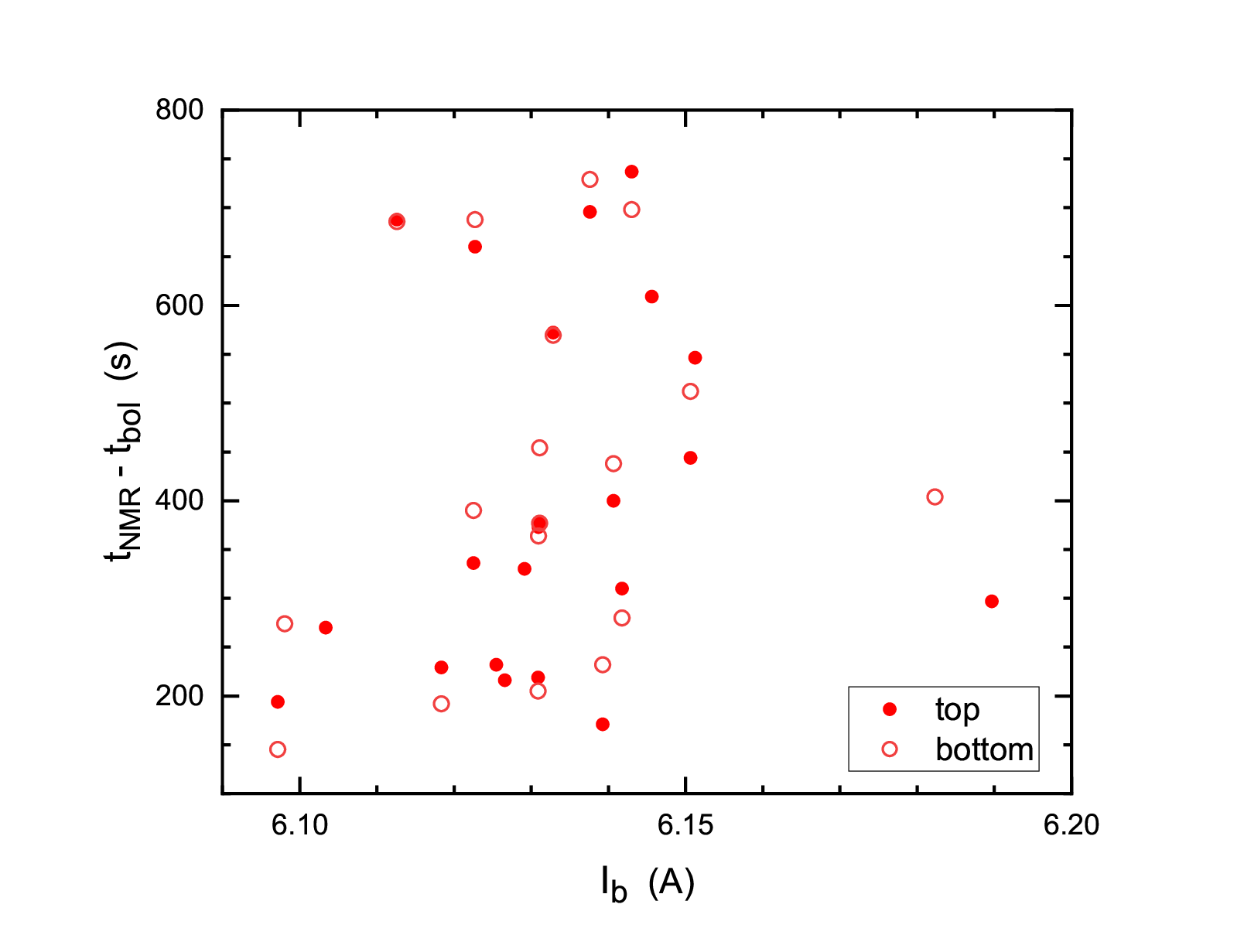}}
\caption{Measured response time $t_\mathrm{NMR} - t_\mathrm{bol}$ plotted versus barrier magnet current $I_\mathrm{b}$ at 29\,bar, $0.20\,T_\mathrm{c}$, and $1.0\,$rad/s. There is no evident correlation between the response time and $I_\mathrm{b}$ (or its sweep rate $dI_\mathrm{b}/dt$). We take this to indicate that the details of the instability event, even in the presence of AB-interface pinning, do not influence the measured response time. %The uncertainty limits represent the maximum error bars, based on the actual disturbances in the bolometer drift. %\textcolor{red}{(Uncertainty limits need to be added!)}
}
\label{ResponseTime_vs_Current}
\end{figure}
%%%%%%%%%%%%%%%%%%%%%%%%%%%%%%%%%%%%%%%%%%%%%%%%%%%%%%%%%%%%%%%%%%%

Fig.~\ref{ResponseTime_vs_Current} shows the response time, $t_\mathrm{NMR} - t_\mathrm{bol}$, plotted as a function of the value of the magnetic field during the downward field sweep, when the instability takes place (given in terms of the current $I_\mathrm{b}$ in the barrier magnet). The data range from 200\,s to 700\,s, with a fair fraction clustering close to the lower limit. 200\,s agrees with the expected flight time, as discussed below. 200\,s was also observed in one exceptional case measured independently of the bolometer signal, when $I_\mathrm{b}$ was kept constant at 6.20\,A and the instability was triggered suddenly by increasing $\Omega$ in 10\,s from 1.004 to 1.044\,rad/s.

\begin{figure}[tt!]
 \centerline{
  \includegraphics[width=0.97\linewidth]{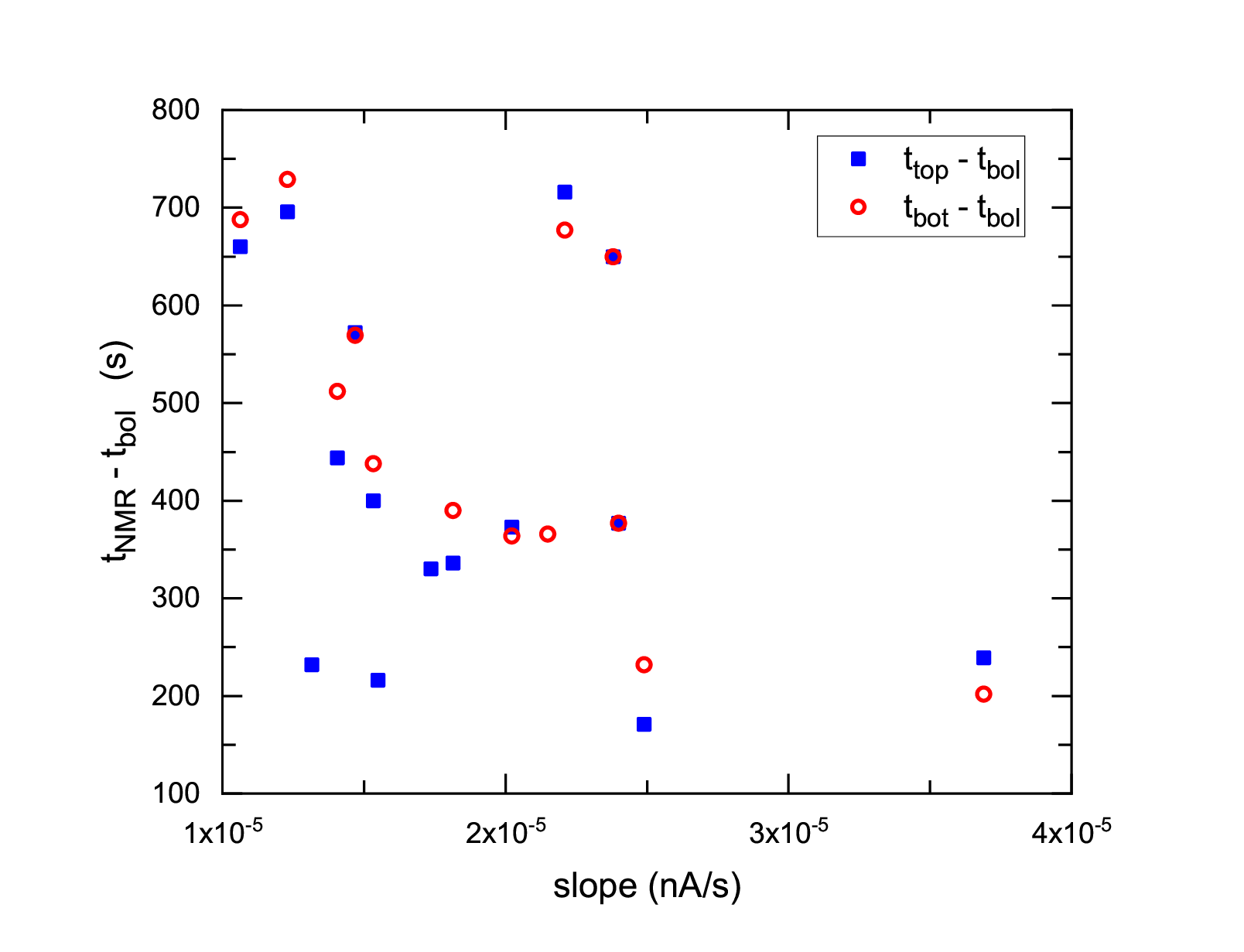}}
\caption{Response time $t_\mathrm{NMR} - t_\mathrm{bol}$ plotted as a function of the slope of the bolometer signal (in arbitrary units) during turbulent vortex front propagation. The expected relation with 1/slope dependence, which connects the propagation time to turbulent heating, is not convincingly illustrated, but rather the plot exemplifies an uncontrolled scatter in the measurement.
%\textcolor{red}{(Fitted curve missing!)}
}
\label{SlopeDependence}
\end{figure}

The long response times in Fig.~\ref{ResponseTime_vs_Current} and the large width of the response time distribution are unexpected new features. To search for their origin, we also checked the response time as a function of the sweep rate, $d\,I_\mathrm{b}/dt$, which varies from $- 1.3 \cdot 10^{-5}$\,A/s to $-1.8 \cdot 10^{-3}$\,A/s. On the basis of these two plots, a correlation between the response time and the destabilizing field value or its sweep rate appear unlikely. In Fig.~\ref{SlopeDependence} we check the response time plotted as a function of the slope of the linear increase in the bolometer signal (dashed line starting from $t_\mathrm{bol}$ in Fig.~\ref{Bolometer}). The dissipation at constant rate by the turbulent front propagating at constant velocity amounts roughly to a 1/slope dependence in this plot. Unfortunately, the scatter of the two readings, of $t_\mathrm{NMR} - t_\mathrm{bol}$ and the slope of the bolometer signal, does neither support or exclude this dependence.

 The KH trigger starts a sequence of different processes of which only the slow events influence noticeably the measured response time $t_\mathrm{NMR} - t_\mathrm{bol}$. The spread of the data to long response times we assume to be a associated with the process with which the KH instability is triggered, namely by sweeping down the barrier field. An entirely different result was obtained in Ref.~\cite{LowT-Front}.

 It was there discovered that a consistent reproducible flight time of the turbulent front was observed down to $0.14\,T_\mathrm{c}$ in a measuring setup where all obstacles and apertures were removed from the long quartz tube. The tube was filled with only B phase (at $I_\mathrm{b}=0$) and carefully annealed from remanent vortices. It could then be rapidly accelerated to rotation, from rest to $\Omega =\;$constant $(\lesssim 1.5\,$rad/s), so that vortex propagation started from the heat exchange sinter at the bottom of the tube. The sinter cannot be cleaned from trapped remnants and provides an instant inexhaustible source of vortices. The propagation of the turbulent vortex front is clocked from its transit time between the two NMR detector coils. In this case all initial processes involving the formation of the turbulent front are reliably excluded from the measured flight time, as even the lower NMR coil is far away from the heat exchange sinter. Reproducible short flight times were measured consistently in this setup.

%%%%%%%%%%%%%%%%%%%%%%%%%%%%%%%%%%%%%%%%%%%%%%%%%%%%%%%%%%%%%%%%%%%
%%%%%%%%%%%%%%%%%%%%%%%%%%%%%%%%%%%%%%%%%%%%%%%%%%%%%%%%%%%%%%%%%%%

The reason for the unexpected extended tail of the data to long response times in Fig.~\ref{ResponseTime_vs_Current} is not clear. One new feature of the present measurements is the destabilizing method where one sweeps down slowly $I_\mathrm{b}$. A characteristic of this trigger is a large variability and deficit in the number of vortices which are initially formed and then contribute to the turbulent vortex front propagation. The thermal calibration of the bolometer reading in Refs.~\cite{Decoupling,ThermalDetection} indicated that the number of vortices responsible for the dissipation during the front propagation is $\lesssim 0.4$ of the equilibrium vortex number. Thus fewer vortices are involved in the turbulent expansion and their number is not increasing during the propagation, owing to the weak vortex-surface interaction and the reduced angular momentum decay of the rotating vortex front. The vortex number starts to grow only later during the slow laminar approach towards the equilibrium state of solid-body rotation \cite{Decoupling}. The estimate from Ref.~\cite{LowT-Front} for the present setup is a flight time $t_\mathrm{NMR} - t_\mathrm{bol} \sim 200$\,s at $0.2\,T_\mathrm{c}$ and $\Omega \approx 1\,$rad/s, in agreement with the lower limit of the data in Fig.~\ref{ResponseTime_vs_Current}. As seen in  Fig.~\ref{Bolometer}, initial processes, such as the formation time of the KH instability or the duration of the turbulent burst which results in the formation of the propagating front, seem to be absent and occur too rapidly to be reproduced by the present bolometer response \cite{ThermalDetection}.

\section{Vortex formation in B $\rightarrow$ A transition}
\label{B->A transition}

A second mechanism for triggering vortex formation in vortex-free B phase counterflow, in a manner similar to the KH instability, is to create the A phase barrier layer by sweeping $I_\mathrm{b}$ upward and triggering the formation of the B $\rightarrow$ A transition at constant rotation and temperature. Instead of sweeping down $I_\mathrm{b}$ to destabilize an existing AB interface in a KH instability, in the upward $I_\mathrm{b}$ sweep the B $\rightarrow$ A transition generates in rapid succession a thin A phase layer, its AB interfaces, and doubly quantized vortex lines in the A-phase. These events simultaneously also start the vortex formation in all of the B phase volume (Fig.~\ref{Bolometer}).

In the starting situation the entire sample cylinder is filled with B-phase vortex-free counterflow, while $I_\mathrm{b}$ is below the value corresponding to $H_\mathrm{AB}$. When $I_\mathrm{b}$ is slowly swept upwards, until the B $\rightarrow$ A transition is observed, a complex sequence of non-equilibrium processes follows. In the final state a uniform narrow A-phase layer separates the two B-phase sections, with all these three sections being filled with vortices close to their equilibrium state. Initially, in rapid succession the A phase and its vortices are formed, since A phase would not be stable without its vortices. Presumably the newly forming A phase remains unstable over much of the cylinder's cross section, until the local counterflow velocities at the AB interfaces on the B-phase side have dropped to sufficiently low value. One could assume that it is emission of vortex rings from the unstable newly forming AB interfaces which starts the turbulence in the B-phase sections.

Let us first compare the B $\rightarrow$ A transition to the reverse case of an A $\rightarrow$ B transition, when the magnetic field is slowly swept downwards past $H_\mathrm{AB}$. Such measurements have established that the final state of the A phase layer before its complete disappearance is an annular ring coating the cylinder wall, with B phase in the center \cite{BarrierMagnet}. This configuration of the AB interface follows the calculated distribution of the barrier field, which is of solenoidal type and increases on moving radially away from the central axis. Experimentally its formation is proven by the observation that the two B-phase sections become interconnected when the AB interface shrinks to an annular ring: if originally one of the B-phase sections is prepared to house vortex-free counterflow and the other the equilibrium vortex state, then the appearance of the hole in the interface allows vortices to fill both sections. In both of these two different configurations of the AB interface the KH instability can be triggered and measured. It is found that the change to the annular interface geometry is marked by a $\sim 10\,$\% drop in $\Omega_\mathrm{c}$ \cite{BarrierMagnet}.

In contrast, in a B $\rightarrow$ A transition, where $I_\mathrm{b}$ is slowly swept upwards, the transition is slightly superheated and A-phase formation does not start smoothly from a localized spot on the rim of the sample cylinder. Instead, the A phase first appears as a complete layer with no central hole. This is proven by the higher $\Omega_\mathrm{c}$  of the KH instability in this initial state. Also, in this situation right after the B $\rightarrow$ A transition the KH instability occurs independently at the two AB interfaces, since one B phase section can be filled with vortices while the other one might still remain in the vortex-free state, demonstrating that the two B-phase sections are not interconnected. Thus, the dynamics of B-phase vortex formation in a KH instability and in a B $\rightarrow$ A transition are expected to display differences.  In the latter case the vortex formation processes are interconnected in both B phase sections in the early phase before the A phase layer reaches a  stable state.

At $0.2\,T_\mathrm{c}$, vortex formation by the B $\rightarrow$ A transition is turbulent and proceeds via vortex ring emission, both upwards and downwards from the newly forming A-phase layer. Small vortex rings travel fast and they may cover all or part of the distance from the A phase layer to the NMR coil. Vortex front propagation may then start in one or more places along the rotating cylinder. Thus the measured time for the vortex front to arrive at the NMR coil can be very short. But another extreme possibility is that the B-phase section is showered with vortex rings so that turbulent vortex formation and thereafter the buildup of rectilinear lines proceeds simultaneously in several places along the B-phase column. In this case the NMR measurement might only display a slow laminar buildup of the vortex state. As displayed by the bolometer signal in Fig.~\ref{Bolometer}, the complete buildup of the equilibrium vortex state requires up to $\sim 1500$\,s at $0.2\,T_\mathrm{c}$.

When the B $\rightarrow$ A transition happens, the bolometer reading registers a prominent cooling spike, caused by the latent heat absorbed in forming the A-phase (Fig.~\ref{Bolometer}). The cooling is also recognized by the NMR measurement if a temperature sensitive feature of the B phase NMR line shape is continuously monitored. Subsequently the bolometer signal continues to evolve, as seen \textit{eg.} in Fig.~\ref{Bolometer}, so that the heating from the turbulent front propagation extrapolates to the start of the cooling spike. If one is not interested in B-phase vortex formation, but only in the critical values of the B $\rightarrow$ A transition, then the starting situation can equally well be the equilibrium B-phase vortex state (inset in Fig.~\ref{Bolometer}). Such a measurement can be carried out much faster than one where the rotating B-phase vortex-free counterflow state has to be prepared.

We note that six B $\rightarrow$ A transitions were analyzed: three were started with B-phase vortex-free counterflow in the sample container and the remaining three with the B phase equilibrium vortex state. The critical current of the B $\rightarrow$ A transition was measured to be $I_\mathrm{b,AB} =$ (5.94---6.01)\,A, while the rotation was $\Omega =$ (0.82---1.20)\,rad/s and the temperature $0.187\,T_\mathrm{c}$ at 29.0\,bar pressure. Using $H_\mathrm{AB}(T,P)$ values from Ref.~\cite{3HeCalculator}, this yields for the six transitions on average a field/current ratio $[H / I_\mathrm{b}]_{B \rightarrow A} = (0.0928 \pm 0.0006)$\,T/A \cite{Correction}. There are no systematic differences in the results between measurements which start from the counterflow or from the equilibrium vortex states. The previous calibration of our barrier magnet \cite{BarrierMagnet}, which has been used in the analysis of all KH measurements including the present, is 0.0942\,T/A and was recorded for B $\rightarrow$ A transitions at a higher temperature of $0.666\,T_\mathrm{c}$. Slight superheating of the AB transition and/or disagreement in  temperature calibrations might explain the difference. The B $\rightarrow$ A transition results are thus as expected and their reproducibility is within the experimental precision.

Our three examples of B $\rightarrow$ A transitions starting from a vortex-free counterflow state displayed differing behaviour in the formation of the equilibrium vortex state and the temporal response, but in each of these cases the evolution was the same in both the top and bottom  sections of the sample container. In two examples the bolometer signals from the turbulent vortex front propagation were of the standard type and gave a response time of 200 s (see Fig.~\ref{Bolometer}). One of these also displayed the NMR signal from vortex front propagation yielding $t_\mathrm{NMR} - t_\mathrm{bol} = 220$s. This case agrees with the flight time $\sim 200\,$s expected after the KH instability (Sec.~\ref{KH_DevelopmentTime}). In the other example the NMR response from vortex front motion was absent, instead a slow laminar buildup of the equilibrium vortex state was recorded which lasted for 1500 s. In contrast to these two examples, the third B $\rightarrow$ A transition measurement did not display a bolometer signal from vortex front propagation but both the bolometer as well as the NMR evidence corresponds to slow laminar formation of the equilibrium vortex state.

The above results on the recovery after a B $\rightarrow$ A transition are consistent with vortex ring emission as the first step in the B-phase vortex formation process. The fact that the recovery behaviour in each of the analyzed events appears to proceed in the same way, when one compares up and down directions, we believe demonstrates that the vortex ring emission process from the newly forming A phase layer operates  in similar fashion in up and down directions. An explanation in terms of vortex ring emission is sufficiently flexible to justify the different responses, but unfortunately, our sample of six  transitions is too small to draw detailed conclusions. Although vortex ring emission appears to be the characteristic feature in the context of the B $\rightarrow$ A transition, it has not been identified in similar manner in our measurements on the KH instability.

\section{Conclusions}
 \label{Conclusions}

 This study examines the KH instability in the temperature regime of ballistic quasiparticle flight, employing both bolometer and NMR measurement for detection. It has been suggested that the bulk-liquid model of the instability, which has proven to explain measurements above $0.4\,T_\mathrm{c}$, might be inadequate in the zero-temperature limit where the coupling of the superfluid component to the external reference frame, mediated by the normal component, is weakened or vanishes altogether. These measurements show that this is not the case: the critical superfluid counterflow velocity triggering the KH instability is found to follow also in the ballistic regime the bulk-liquid model developed for higher temperatures.

 The agreement rests on fitting the measurements  with an AB interfacial surface tension $\sigma_{\mathrm{AB}} (T, P, H) = \sigma_0 (T,P) \; (1-a_\mathrm{H}(P) [H_{\mathrm{AB}} (T,P)]^2)$,  to determine the coefficient $a_\mathrm{H}(P)$ of the first order magnetic-field-dependent correction term. At $P = 29.0$\,bar pressure and $0.2\,T_\mathrm{c}$ this amounts to a suppression $1-a_\mathrm{H}(P) [H_{\mathrm{AB}} (T,P)]^2 \approx 0.38 $. In these conditions the KH instability proves to be a rapid non-equilibrium process with a development time which is too fast to be resolved with the present measuring techniques.

 In addition, the temporal recovery to equilibrium rotation after the KH instability has been studied. Earlier measurements at higher temperatures displayed a predictable regular recovery process. In contrast, at $0.2\,T_\mathrm{c}$ the response times are less reproducible, displaying a wide distribution, which extends upward from the cut-off given by the expected flight time of the turbulent vortex front. The AB interface is here destabilized with a downward current sweep, whereby slight irreproducibility is noticed, consistent with weak AB interface pinning at the smooth-walled quartz sample container. This triggering method is believed to result in more variation of the starting conditions and contributing to the variability in the measured response times.

 Finally, instead of the standard KH measurement where the AB interface is made unstable, we also explored the inverse case where a virgin new layer of A phase is created in the rotating column filled with metastable vortex-free B phase, by sweeping the magnet current upward. Here the recovery behaviour was found to be more variable, the expected turbulent dissipation of $\sim 200\,$s duration was recorded for two examples from three. In the third case the turbulent heating was absent and only a slow laminar process was observed.

 These results can be explained in terms of vortex rings escaping from the location where the first-order B $\rightarrow$ A phase transition takes place in the region of maximum magnetic field. The rapidly moving vortex rings (with a radius comparable to the inter-vortex distance at the constant $\Omega \sim 1\,$rad/s) can start the B-phase turbulent vortex front motion at varying distances from the newly forming A-phase layer and in this way lead to a wide range of apparent flight times for the turbulent front. Here vortex ring formation and their emission from the highly non-equilibrium B $\rightarrow$ A transition layer are part of a complex process, where from vortex-free B-phase counterflow the A phase, its vortices, and the two AB interfaces all have to appear simultaneously. This process has not been studied further.

\textit{\textbf{Acknowledgements}}:--This contribution to the Andreev Memorial Collection is a tribute to Academician Alexander Feodorovich Andreev, a brilliant scientist, our loyal mentor in both scientific and organizational questions, and a personal friend. We are indebted to him for his advice and decades-long support in funding and managing the research on rotating $^3$He superfluids under the umbrella of the Finnish and Russian Science Academies.  The measurements were carried out in the Low Temperature Laboratory, which is part of the OtaNano research infrastructure of the Aalto University.

\vspace{6mm} ${^\ddag}$Correspondence email: matti.krusius@aalto.fi


\begin{thebibliography}{10}
\bibitem{Correction} No correction has been applied for the hysteretic displacement of our measured B $\rightarrow$ A transition above the thermodynamic equilibrium value.
\end{thebibliography}
\end{document}